# Feedback and Timing in a Crowdsourcing Game

Gili Freedman, Sukdith Punjasthitkul, Max Seidman, and Mary Flanagan

Tiltfactor Lab, 245 BFVAC, HB 6194 Dartmouth College, Hanover, NH 03755
contact@tiltfactor.org

**Abstract**

The present research examines two problems inherent to the creation of crowdsourcing games: how to give feedback when the right answer is not always known by the game and how much time to give players without sacrificing data quality. Taken together, the present research provides an important first step in considering how to create fun, challenging crowdsourcing games that generate quality data.

## Introduction

Crowdsourcing games have the potential to appeal to wide audiences and generate data in a fun way (Law & von Ahn 2009; von Ahn & Dabbish 2008), but there are important challenges inherent to creating games that rely on human computation. One challenge is that of feedback. Games generally involve feedback: players know when they have performed correctly or incorrectly based on how they are rewarded (Hunicke, LeBlanc, & Zubeck 2004). However, in a crowdsourcing game, the game creators typically do not know the correct answer ahead of time, and rewards are thus not always perfectly matched to player accuracy. Therefore, the present study examined how varying levels of feedback influenced players' perceptions of the game. A second problem with crowdsourcing games is that of timing. Games often use time limits to increase the challenge (Hunicke et al. 2004), but crowdsourcing games require the best quality answer, which may require providing more time. Therefore, we examined how varying levels of time influenced both player perception of the game and quality of the data.

## Methods

### Feedback Study

One hundred fifty (64 female; $M_{age}$ = 34.88, $SD_{age}$ = 10.98) Amazon Mechanical Turk (MTurk) workers completed this study and were compensated $1 USD.

In both the Feedback Study and the Timing Study (see below), participants played a sorting game in which they viewed 20 images and had to indicate whether the images contained an element described by a text keyword. After indicating their choice, they were given audio (correct "bing" vs. incorrect "buzz") and visual feedback (a green vs. red button outline; score went up as well in correct) as to whether they were correct. In the Feedback Study, participants were randomly assigned to one of three conditions: 50% correct feedback (i.e., half of the time if they got the correct answer they were told it was incorrect and vice versa), 90% correct feedback, or 100% correct feedback. Then participants completed a questionnaire.

### Timing Study

Two hundred fourteen (101 female, 1 did not report gender; $M_{age}$ = 34.84, $SD_{age}$ = 11.19) Amazon MTurk workers completed this study and were compensated $1.

Players in the Timing Study saw the same set of images in the same sequence as in the Feedback Study and were only given correct feedback. However, the amount of time they had to answer each question varied. Participants were randomly assigned to 2, 4, or 10 seconds, or unlimited time. After sorting images, participants completed the post-game questionnaire.

## Results

### Feedback Study

The level of correct feedback did not significantly influence participants' performance quality: $F(2, 147) = 1.36$, $p = .26$. However, the level of correct feedback did influence perceptions of how complicated ($F(2, 147) = 15.48$, $p < .001$) and challenging ($F(2, 147) = 14.95$, $p < .001$) the game was as well as whether they felt they could master the game ($F(2, 147) = 5.78$, $p = .004$). Specifically, participants in 50% correct feedback found the game more complicated ($M = 4.50$, $SD = 2.30$) than participants in 90% correct feedback ($M = 2.98$, $SD = 1.70$; $t(94) = 3.71$, $p <$

.001) or 100% correct feedback ($M = 2.48$, $SD = 1.58$; $t(98) = 5.19$, $p < .001$). There was no difference in participants' perceptions of how complicated the game was between those in 90% vs. 100% correct feedback ($t(102) = 1.55$, $p = .12$). A similar pattern was found for how challenging the game was perceived to be. Participants in 50% correct feedback found the game more challenging ($M = 4.33$, $SD = 2.29$) than participants in 90% correct feedback ($M = 2.90$, $SD = 1.46$; $t(94) = 3.67$, $p < .001$) or 100% correct feedback ($M = 2.44$, $SD = 1.50$; $t(98) = 4.92$, $p < .001$). There was no difference in participants' perceptions of how challenging the game was between 90% vs. 100% correct feedback ($t(102) = 1.57$, $p = .12$). Participants in 50% correct feedback felt that they were less likely to master the game ($M = 5.93$, $SD = 3.01$) than participants in 90% correct feedback ($M = 7.14$, $SD = 2.40$; $t(94) = 2.18$, $p = .032$) or 100% correct feedback ($M = 7.65$, $SD = 2.26$; $t(98) = 3.24$, $p = .002$). There was no difference in participants' perceptions of how likely they were to master the game between those who received 90% vs. 100% correct feedback ($t(102) = 1.11$, $p = .27$).

Feedback did not impact how interesting ($F(2, 147) = 1.41$, $p = .25$) or rewarding ($F(2, 147) = 2.17$, $p = .12$) the game was or how likely participants were to indicate they would play again ($F(2, 147) = 5.49$, $p = .45$).

**Timing Study**

The amount of time allotted significantly influenced participants' performance quality: $F(3, 210) = 21.78$, $p < .001$. Specifically, participants provided more correct answers in Unlimited Time ($M = 19.26$, $SD = 1.31$) than 2 seconds ($M = 16.20$, $SD = 3.14$; $t(102) = 6.40$, $p < .001$) or 4 seconds ($M = 17.84$, $SD = 2.83$; $t(103) = 3.25$, $p = .002$). Furthermore, participants answered correctly more often in 10 seconds ($M = 19.33$, $SD = 1.22$) than in 2 seconds ($t(107) = 6.88$, $p < .001$) or 4 seconds ($t(108) = 3.59$, $p = .001$). Finally, participants answered correctly more often in 4 seconds than in 2 seconds ($t(107) = 2.85$, $p = .005$).

Time also influenced perceptions of how complicated ($F(3, 210) = 2.98$, $p = .03$) and challenging ($F(3, 210) = 5.06$, $p = .002$) the game was as well as whether they felt they could master the game ($F(3, 210) = 2.66$, $p = .05$). Specifically, participants in 2 seconds found the game more complicated ($M = 3.44$, $SD = 1.80$) than participants in Unlimited Time ($M = 2.58$, $SD = 1.44$; $t(102) = 2.69$, $p = .008$) or 10 seconds ($M = 2.76$, $SD = 1.49$; $t(107) = 2.15$, $p = .03$). A similar pattern was found for perceptions of challenge: participants in 2 seconds found the game more challenging ($M = 3.69$, $SD = 1.97$) than participants in Unlimited Time ($M = 2.54$, $SD = 1.50$; $t(102) = 3.32$, $p = .001$) or 10 seconds ($M = 2.71$, $SD = 1.47$; $t(107) = 2.93$, $p = .004$). Additionally, participants in 2 seconds found the game more challenging than participants in 4 seconds ($M = 2.89$, $SD = 1.57$; $t(107) = 2.33$, $p = .02$). Furthermore, participants in 2 seconds ($M = 7.22$, $SD = 2.43$) and 4 seconds ($M = 7.22$, $SD = 2.34$) felt that they were less likely to master the game than participants in 10 seconds ($M = 8.24$, $SD = 1.73$; $t(107) = 2.51$, $p = .01$; $t(108) = 2.60$, $p = .01$, respectively.

Timing did not impact how interesting ($F(3, 210) = 1.23$, $p = .30$) or rewarding ($F(3, 210) = 1.41$, $p = .24$) the game was or how likely participants were to indicate they would play again ($F(3, 210) = .12$, $p = .95$).

## Discussion

The present research provided two important pieces of information for people interested in creating games for crowdsourcing information. First, providing some false feedback does not negatively impact the quality of the data or the desire to play the game again, but it does make the game feel more challenging, complicated, and difficult to master. One limitation of the present study was that the task was fairly easy, and the effect of false feedback may have been hidden by a ceiling effect in the number of correct responses. Second, providing a limited time frame can negatively impact the quality of the data, and it makes the game feel more challenging, complicated, and difficult to master. Future research can build on these important first steps by finding the optimal time allotment for given tasks that make the task challenging without decreasing data quality. Importantly, feedback and timing both do not impact how interesting or rewarding games are perceived to be. For the field of game design, this is an important contribution, for games can therefore captivate and feel rewarding outside of feedback and temporal structures. Furthermore, future research can continue to examine the role of feedback on player performance in crowdsourcing games.

## Acknowledgments

The present research was funded by the National Endowment for the Humanities [HK-50021-12]. We also gratefully thank all of our participants and student assistants.